# Fiber as a temperature sensor with portable Correlation-OTDR as interrogator

Florian Azendorf
Advanced Technology
ADVA Optical Networking SE
Meiningen, Germany
fazendorf@adva.com

Annika Dochhan
Advanced Technology
ADVA Optical Networking SE
Meiningen, Germany
adochhan@adva.com

Florian Spinty
Advanced Technology
ADVA Optical Networking SE
Meiningen, Germany
fspinty@adva.com

Mirko Lawin
Advanced Technology
ADVA Optical Networking SE
Meiningen, Germany
mlawin@adva.com

Bernhard Schmauss
Optical Radio Technology and Photonics
Friedrich-Alexander-University Erlangen/Nürnberg
Erlangen, Germany
bernhard.schmauss@fau.de

Michael Eiselt
Advanced Technology
ADVA Optical Networking SE
Meiningen, Germany
meiselt@adva.com

*Abstract*—In this paper, we report on the integration of a Correlation-OTDR into a portable unit (30 cm x 30 cm), based on a multi-processor system on chip (MPSoC) and a small formfactor pluggable (SFP) 2.5G transceiver. Going beyond telecommunication applications, this system is demonstrated for temperature measurements, based on the change of propagation delay with temperature in a short section of optical fiber. The temperature measurement accuracy is investigated as a function of the fiber length from 4 to 25 meters.

## I. Introduction

For the next generation networks, the propagation delay of an optical signal in the fiber becomes a critical parameter. Synchronization protocols such as Precision Time Protocol (PTP) require a symmetric propagation delay between master and slave clock and vice versa. For current synchronization requirements, an asymmetry on the order of tens of nanoseconds is sufficient. Increasing synchronization accuracy, however, will reduce this value to a nanosecond or even less. The propagation delay in the fiber is affected by fluctuations in the environmental temperature. In a previous work we characterized fibers in an underground cable with a length of 8 km and measured a change in group delay of 800 ps for all fibers over the first 6 days in summer [1]. In another work we characterized three different jumper cables and showed that the temperature effect on the propagation delay is partly based on the outer jackets [2]. The temperature impact can be compensated for, but it must be monitored. For this application, the Correlation-Optical Time Domain Reflectometry (C-OTDR) method was developed to measure the group delay of the optical signal. The investigations in the previous work were performed with a laboratory setup including stationary equipment. To enable mobile measurements in the field, we now report the integration of the measurement setup in a multi-processor system-on-chip (MPSoC) in a field programmable gate array (FPGA). With the high-speed data interface, it is feasible to generate and to receive 10 Gbit/s signals. In this paper, we compare the laboratory setup with the portable setup.

## II. Measurement Principle

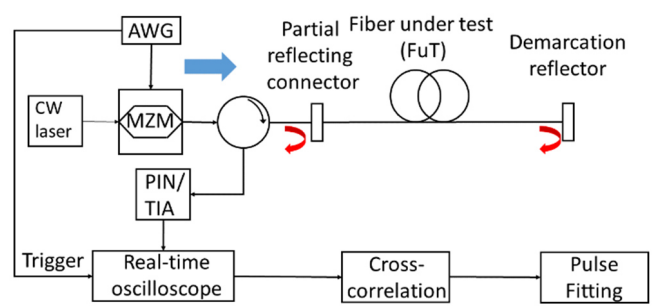

Fig. 1. Laboratory setup of the Correlation-OTDR.

In Fig. 1 the typical laboratory setup for the Correlation-OTDR is shown. The electrical signal, which contains a burst with defined length and a series of zeros, is generated with an arbitrary waveform generator (AWG). This electrical signal is modulated onto a continuous wave signal with a particular wavelength with a Mach-Zehnder modulator. The signal is fed into an optical circulator and reflected at a partially reflecting connector to obtain a reference. This reference is needed to calculate the absolute latency of the fiber under test (FuT). At the fiber output an open connector was used to obtain a reflection from the end. The reflected signals are received with a PIN + TIA combination and recorded with a real-time oscilloscope. To improve the signal to noise ratio (SNR) one thousand individual signals were recorded. In the laboratory setup the signals were sampled at 50 GS/s. Afterwards, the signals were processed offline. The first signal processing step is the cross-correlation of the received signal and the transmitted burst. Thus, we obtain sharp reflection peaks. A Gaussian function was fitted to the reflection peaks. Through the fitting we achieve a precision better than one sample period.

## III. Portable Correlation-OTDR

The setup which was used in previous works [3] used expensive and bulky equipment like the AWG and the real-time oscilloscope. Measurements were therefore confined to the laboratory. To enable mobile measurements, the C-OTDR technique was integrated into an MPSoC in an FPGA for signal generation, recording, and processing. For cost reduction, the discrete optical front-end components were substituted with an SFP transceiver. The probing signal is generated with a data rate of 2.5 Gbit/s. Each frame contains a 512 Bit Golay sequence plus a fill pattern. The frame is longer than the round-trip time in the FuT to obtain unambiguous reflections. The reflected signal was sampled with 10 GS/s, thus 4-fold oversampling of the transmitted bit rate. The receiver side consists of an avalanche photodiode (APD) with a transimpedance amplifier and a limiting amplifier. The 7-bit analog to digital converter of the real-time oscilloscope was substituted by 1-bit slicer as a combination of the limiting amplifier in the SFP and the high-speed data input on the FPGA. Multiple received signal traces were averaged to increase the amplitude resolution. The bit and frame clocks of transmitter and receiver were phase aligned to avoid jitter during the accumulation of the bits. The incoming sampled and sliced signal is transferred into memory and accumulated. After summing all received traces, the sums for each time slot get transferred to the CPU in the MPSoC, where the data processing is performed.

## IV. MEASUREMENTS AND RESULTS

The propagation delay in the optical fiber is sensitive to fluctuations in the ambient temperature. A typical unjacketed single mode optical fiber with coating has a temperature coefficient of delay (TCD) of 7.49 ppm/K [4]. This value is a sum of the thermal expansion of the fiber length and the temperature induced change of the refractive index.

### A. Group delay variation over 40K

In the first experiment we want to measure this TCD for bare fibers with different lengths. Therefore, we placed three different bare fibers with a length of 4 m, 10 m, and 25 m in a temperature-controlled cabinet. The three fibers were cascaded in the order 4 m, 10 m, and 25 m. To obtain reflections, partial reflecting connectors were used between the fiber sections. For a temperature change of 10 K, we expected a change of the round-trip propagation delay of approximately 3 ps, 7.5 ps, and 18.7 ps for 4 m, 10 m, and 25 m fiber, respectively. The temperature was increased in 10 K steps in a range of 10 °C to 50 °C. After the oven temperature settled, ten measurements were performed for each temperature. In Fig 2 the temperature change over time is shown.

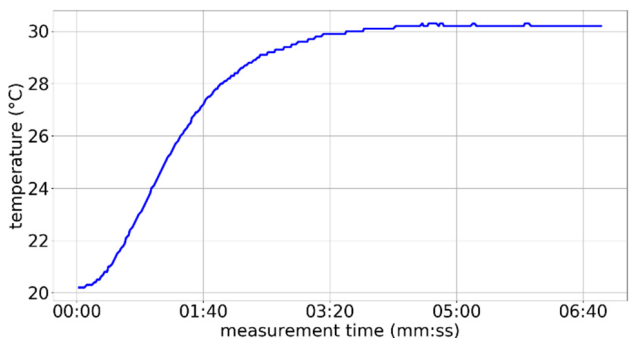

Fig. 2. Oven temperature over time.

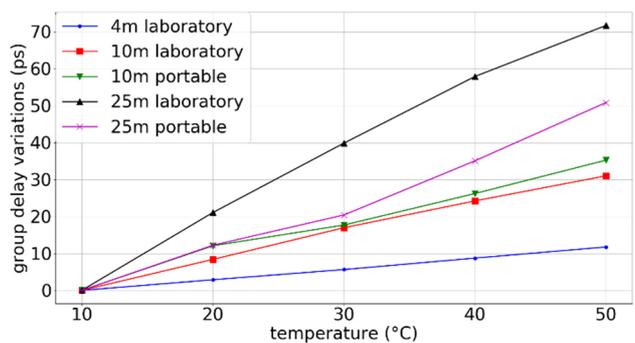

Fig. 3 Propagation delay changes over a temperature range of 40 K.

Fig. 3 shows the group delay increase from 10 °C to 50 °C for the different bare fiber lengths. The results show the expected linear slope of the group delay as a function of temperature. With the laboratory setup we achieved the expected change of the group delay with small fluctuations of approximately 0.1 ps. With the portable setup the change of the group delay of the 4m fiber was not measurable due to large variations and measurement errors. The values of the 10 m fiber were similar for the laboratory and the portable setup. The 25 m fiber measured with the portable C-OTDR showed larger deviations from the laboratory setup results. The resulting error of these deviations is about 40 ppm of the fiber propagation delay. So far, no explanation of these errors has been found. With the measured values we obtained for all fibers, except the 25 m fiber, a TCD of approximately 7 ppm/K.

### B. Fast measurements of temperature variations

One advantage of the portable C-OTDR over the laboratory setup is the fast signal processing such that measurements can be performed in intervals of 2 seconds, even with averaging over 4000 traces per measurement. In the second experiment the bare fibers were placed again in the temperature-controlled cabinet, and the temperature was varied between 25 °C and 35 °C. Using the portable setup, the group delay change in the cabinet during the heating and cooling process was measured. It was not possible to characterize the 4 m fiber, because the propagation delay changes with only 300 fs per Kelvin and the time resolution of the portable C-OTDR is not sufficient to measure such small fluctuations. The results of the measurements for the 10-m fiber are shown in Fig. 4 together with the temperature measured by a temperature sensor inside the chamber, which was interrogated with a digital multimeter. To compare the measured temperature and the measured group delay we converted the group delay to temperature, by using the calculated TDC of 7 ppm/K of the first experiment

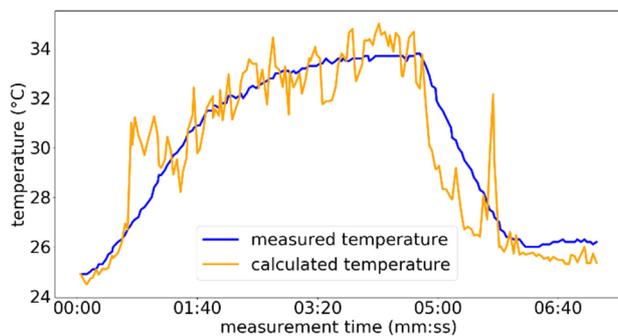
Fig. 4. Temperature measured by thermometer and by group delay method of 10m fiber.

Deviations between the measured temperature of the fiber and the oven can be observed. During the heating period, the error of the standard deviation was 2.86 K and can be explained by measurement errors of the system. During the cooling down period, the fiber appeared to cool down faster that the air measured with the temperature sensor. We assume the faster cooling is caused by the ovens air stream, which was directed at the fiber.

## V. Conclusion

In this paper we present the integration of a C-OTDR into an MPSoC in an FPGA. The comparison of the laboratory setup and portable setup showed differences between the setups. Nevertheless, we showed that a 10-meter section of fiber can be used to measure the temperature with an accuracy of approximately 1K with the portable setup. The portable C-OTDR was also used to measure the temperature variations in 2-second intervals during the heating and cooling process of a 10-meter fiber section. Further improvement of the portable measurement setup will help to reduce the measurement errors.

## Acknowledgment

This project has received funding from the European Union´s Horizon 2020 research and innovation programme under grant agreement No 762055 (blueSpace Project)